\documentclass[10pt]{llncs}
\usepackage[pdftex]{graphicx}

\usepackage{amsmath}
\usepackage{subfigure}
\usepackage{upgreek}
\usepackage{times}
\usepackage{multirow}
\usepackage{array}

\begin{document}

\spnewtheorem{observation}{Observation}{\bf}{\it}

%
\titlerunning{Bisection (Band)Width of Product Networks with Application to Data Centers}

\title{Bisection (Band)Width of Product Networks with Application to Data Centers\thanks{This research was supported in part by the Comunidad de Madrid grant S2009TIC-1692, Spanish MICINN grant TEC2011-29688-C02-01, and National Natural Science Foundation of China grant 61020106002.}}
\titlerunning{Bisection (Band)Width of Product Networks with Application to Data Centers}

\author{Jordi Arjona Aroca\inst{1}\inst{2} \and Antonio Fern\'andez Anta\inst{2}}

\institute{Institute IMDEA Networks, Madrid, Spain
\and
Universidad Carlos III de Madrid, Madrid, Spain}

\maketitle

\begin{abstract}
The bisection width of interconnection networks has always been important in parallel computing, since it bounds the amount of information that can be moved from one side of a network to another, i.e., the bisection bandwidth. Finding its exact value has proven to be challenging for some network families.
For instance, the problem of finding the exact bisection width of the multidimensional torus was posed by Leighton
and has remained open for almost $20$ years.
In this paper we provide the exact value of the bisection width of the torus, as well as of several $d$-dimensional classical parallel topologies that can be obtained by the application of the Cartesian product of graphs. To do so, we first provide two general results that
allow to obtain upper and lower bounds on the bisection width of a product graph as a function of some properties of its factor graphs. We also apply these results to obtain bounds for the bisection bandwidth of a $d$-dimensional BCube network, a recently proposed topology for data centers.

\end{abstract}



\DeclareGraphicsExtensions{.pdf}

\section{Introduction}

The bisection width and the bisection bandwidth of interconnection networks have always been two important parameters of a network.
The first one reflects the smallest number of links which have to be removed to split the network in two equal parts, while the second one bounds the amount of data that can be moved between these parts. In general, both values are derivable one from the other, which is the reason why
most previous work has been devoted to only one of then (in particular, the bisection width).

The bisection width has been a typical goodness parameter to evaluate and compare interconnection networks for parallel architectures \cite{leighton,duato,dallyTowles}.
This interest has been transferred to the Network-On-Chip topologies, as the natural successors of the parallel architectures of the 90's \cite{odi,empirical,torusbased,NOCsurvey}. 
The bisection (band)width is also nowadays being used as a reference parameter on the analysis of the latest topologies that are being deployed in data centers. This can be seen in recent papers which propose new topologies, like BCube\cite{bcube} or DCell \cite{dcell}. The bisection (band)width is used to compare these new topologies with classical topologies, like grids, tori, and hypercubes, or with other datacenter topologies, like trees and fat trees.

Finding the exact value of the bisection width is hard in general.
Computing it has proven to be challenging even for very simple families of graphs.
For instance, the problem of finding the exact bisection width of the multidimensional torus was posed by Leighton \cite[Problem $1.281$]{leighton} and has remained open for almost $20$ years.
One general family of interconnection networks, of which the torus is a subfamily, is the family of product networks. The topology of
these networks is obtained by combining factor graphs with the Cartesian product operator. This technique allows to build large networks
from the smaller factor networks. Many popular interconnection networks are instances of product networks, like the grid and the hypercube.
In this paper we derive techniques to bound the bisection width of product networks, and apply these techniques to obtain the bisection width
of some product network families.

\subsection{Related work}

To our knowledge, Youssef \cite{Youssef91,Youssef95} was among the first to explore the properties of product networks as a family. He presented the idea of working with product networks as a divide-and-conquer problem, obtaining important properties of a product network in terms of the properties of its factor graphs.

The bisection width of arrays and tori was explored by Dally \cite{dally} and Leighton \cite{leighton} in the early $90$s, presenting exact results for these networks when the number of nodes per dimension was even. The case when there are odd number of nodes per dimension was left open. Rolim et al. \cite{rolim95} gave the exact values for the bisection width of $2$ and $3$-dimensional grids and tori, but left open the question for longer number of dimensions.


For the special case in which all the factors are isomorphic, Efe and Fern\'andez \cite{efernan3} provided a lower bound on the bisection width of a product graph as a function of a new parameter of a factor network they defined, the maximal congestion.
Nakano \cite{nakano} presented the exact value of the bisection width for the Cartesian product of isomorphic paths and cliques (i.e., square grids and Hamming graphs). If the factor graphs have $k$ nodes, he proved that the $d$-dimensional square grid has bisection width $k^{d-1}$ when $k$ is even, and $\frac{(k^d-1)}{(k-1)}$ when $k$ is odd. Similarly, the square Hamming graph has bisection width $k^{d+1}$ when $k$ is even, and $(k+1)\frac{(k^d-1)}{4}$ when $k$ is odd. The exact bisection width of the $d$-dimensional square grid was found independently by Efe and Feng\cite{efe}.

For the present paper it is very relevant the work of Azizoglu and Egecioglu. In \cite{cylinders} and \cite{azizo} they studied the relationship between the isoperimetric number and the bisection width of different product networks. In the former paper, they find the exact value of the bisection width of the cylinders (products of paths and rings) with even number of nodes in its largest dimension. In the latter reference
they found the exact bisection width of the grid $A^{(d)}_{k_1,k_2,\ldots, k_d}$, with $k_i$ nodes along dimension $i$, and where $k_1\geq k_2 \geq \ldots \geq k_d$. The value of this bisection width is $BW(A^{(d)}_{k_1,k_2,\ldots, k_d})=\sum^\alpha_{i=1}C_i$,
where $\alpha$ is the smallest index for which $k_i$ is even ($\alpha=d$ if no index is even), and $C_i=\prod_{j=i+1}^d{k_j}$.


%

\subsection{Contributions}

In this paper we present two theorems that allow to derive lower and upper bounds on the bisection width of a product network as a function of some
simple parameters of its factor graphs. Then, we apply these results to obtain the exact value of the bisection width for several families of
product networks. The families presented are of interest because they have been proposed as interconnection networks for parallel architectures,
but their bisection width has never been derived exactly.

One of the most interesting contribution of this paper is the exact value of the bisection width of the torus, since, as mentioned before, this problem has been open for almost $20$ years. We find here that the exact value of the bisection width of a $d$-dimensional torus $T^{(d)}_{k_1,k_2,\ldots, k_d}$, that has $k_i$ nodes along dimension $i$, and where $k_1\geq k_2 \geq \ldots \geq k_d$, is exactly twice the bisection width of the grid of similar
dimensions $A^{(d)}_{k_1,k_2,\ldots, k_d}$. I.e.,
$$
BW(T^{(d)}_{k_1,k_2,\ldots, k_d})=2\sum^\alpha_{i=1}C_i,
$$
where $\alpha$ is the smallest index for which $k_i$ is even ($\alpha=d$ if no index is even), and $C_i=\prod_{j=i+1}^d{k_j}$.
Since this value will appear frequently, we will use the following notation throughout the rest of the paper,
\begin{equation}
\label{eq:psi}
    \Uppsi(\alpha) = \sum_{i=1}^\alpha{C_i} = \sum_{i=1}^\alpha{\prod_{j=i+1}^d{k_j}}.
\end{equation}
Hence, $BW(A^{(d)}_{k_1,\ldots, k_d})=\Uppsi(\alpha)$ and $BW(T^{(d)}_{k_1,\ldots, k_d})=2\Uppsi(\alpha)$.
In addition to the result for the torus, we provide the exact value for the bisection width of products of complete binary trees (CBT) of any size (mesh connected trees \cite{efernan}), products of extended CBT (which are CBT with the leaves connected with a path \cite{efernan}), products of CBT and paths, and products of extended CBT and rings.
To obtain the bisection bandwidth of these networks, we assume that every edge removed by the bisection width is in fact a duplex link with bandwidth of $T$ in each direction. This directly implies that for any of these networks $G$, the bisection bandwidth is computed as $BBW(G)=2T \cdot BW(G)$.

The general upper and lower bound results are also used to derive bounds on the bisection bandwidth of a topology proposed for datacenters, the BCube. A BCube is the Cartesian product of factors networks formed by $k$ nodes connected via a $k$-port switch (where the switch is not considered to be a node). An essential difference of this topology from the previous one is that edges do not connect nodes directly, and the direct
relation between bisection width and bisection bandwidth does not hold anymore. In networks with switches like this one, the switching capacity $s$ of the switch comes into play as well. Since the bisection bandwidth is the parameter of interest in datacenters, we derive bounds on its value for two cases: when the bottleneck for the bisection bandwidth is fully at the switches, and when it is fully  at the links.

Table \ref{tab:tech} summarizes the results for the bisection bandwidth obtained for the different parallel topologies and for BCube. As can be seen there, for the former the values obtained are exact, while for the latter upper and lower bounds do no match exactly. However, they differ by less than a factor of two.

\begin{table*}[tb]
    \caption{Bisection bandwidth of different product networks}
    \label{tab:tech}
    \centering
        \begin{tabular}{|c|c|c|c|c|c|}
                \hline
                        \textbf{Product graph} 	& \multicolumn{2}{c|}{\textbf{Factor graphs}} 	    & \textbf{$\beta(G)$} 	& \textbf{$CC(G)$}	& \textbf{Bisection bandwidth} \\
                \hline
                        Torus &	\multicolumn{2}{c|}{Ring} 	& $1/8$ 	& $2$ 	& $4T\cdot\Uppsi(\alpha)$ \\
                \hline
                        Product of extended CBT &	\multicolumn{2}{c|}{XTs} 	& $1/8$ 	& $2$ 	& $4T\cdot\Uppsi(\alpha)$ \\
                \hline
                        Product of extended CBT \& rings &	\multicolumn{2}{c|}{Rings \& XTs} 	& $1/8$ 	& $2$ 	& $4T\cdot\Uppsi(\alpha)$ \\
                \hline
                        Mesh connected trees &	\multicolumn{2}{c|}{CBT} 	& $1/4$ 	& $1$ 	& $2T\cdot\Uppsi(\alpha)$ \\
                \hline
                        Product of CBT and paths &	\multicolumn{2}{c|}{Paths \& CBTs} 	& $1/4$ 	& $1$ 	& $2T\cdot\Uppsi(\alpha)$ \\
                \hline
                        \multirow{4}{*}{BCube} &	\multirow{2}{*}{Model A} &even	& $\frac{k-1}{k^2}$	& $\frac{k}{2}$ 	& $2T\frac{k^{d+1}}{4(k-1)} \leq BBW(BCA^{(d)}_{k}) \leq 2T\frac{k^d}{2}$ \\
                \cline{3-6}
                                                                       &     & odd	& $\frac{1}{k+1}$	& $\frac{k-1}{2}$ 	& $2T\frac{k+1}{4}\frac{k^d-1}{k-1} \leq BBW(\mathit{BCA}^{(d)}_{k}) \leq 2T\frac{k^d-1}{2}$ \\
                \cline{2-6}
                                              &      \multirow{2}{*}{Model B} &even	& $\frac{k-1}{2k}$ 	& $1$ 	& $   s\frac{k^d}{2(k-1)} \leq BBW(\mathit{BCB}^{(d)}_{k}) \leq s\frac{k^d-1}{k-1}$ \\
                \cline{3-6}
                                                                     &       &odd	& $\frac{k}{2(k+1)}$ 	& $1$ 	& $s\frac{k+1}{2k}\frac{k^d-1}{k-1} \leq BBW(\mathit{BCB}^{(d)}_{k}) \leq s\frac{k^d-1}{k-1}
$ \\
                \hline

        \end{tabular}
\end{table*}

%
%
%

The rest of the paper is organized as follows. Section \ref{sec:defs} presents some basic definitions used in the rest of sections. In Section \ref{sec:bounds} we provide the general results to derive bounds on the bisection bandwidth of product networks. Section \ref{sec:pathtree} and Section \ref{sec:ringxt} present our results for the bisection bandwidth of some classical parallel topologies. Bounds on the bisection bandwidth of the BCube network are presented in Section \ref{sec:bcube}. Finally, in Section~\ref{s-conclusions} we present our conclusions and some open problems.


\section{Definitions}
\label{sec:defs}

\subsection{Graphs and bisections}

In this section we present definitions and notation that will be used along the text. Given a graph\footnote{Unless otherwise stated we will use the terms graph and network indistinctly.} $G$, we denote its sets of vertices and edges as $V(G)$ and $E(G)$, respectively. In some cases, when it is clear from the context, only $V$ or $E$ will be used, omitting the graph $G$. Unless otherwise stated, the graphs considered are undirected.

Given a graph $G$ with $n$ nodes, we use $S(G)$ to denote a subset of $V(G)$ such that $|S(G)| \leq \frac{n}{2}$.
We also use $\partial^G S(G)$ to denote the set of edges connecting $S(G)$ and $V(G) \setminus S(G)$.
Formally, $\partial^G S(G) = \{(u,v)\in E(G):u\in S(G), v\in G\setminus S(G)\}$. The graph $G$ may be omitted from this notation when it is clear from the context.

The main object of this work is to calculate the bisection width and bisection bandwidth of different product networks. These bisections can be defined as follows.

\begin{definition}
The \emph{bisection width} of an $n$-node graph $G$, denoted $BW(G)$, is the smallest number of edges that have to be removed from $G$ to partition it in two halves.
Formally, $BW(G)=\min_{S:|S|=\lfloor \frac{n}{2}\rfloor}{|\partial^G S|}$.
\end{definition}

\begin{definition}
The \emph{bisection bandwidth} of a network $G$, denoted $BBW(G)$, is the minimal amount of traffic which can be transferred between any two halves of the network when its links are transmitting at full speed.
\end{definition}

As mentioned above, unless otherwise stated we assume that all the links in a network $G$ are duplex and have the same capacity $T$ in each direction. Then, we can generally assume that the relation between the bisection bandwidth and the bisection width is $BBW(G)=2T\cdot BW(G)$.


\subsection{Factor and product graphs}

We define first the Cartesian product of graphs.

\begin{definition}
\label{def:cartesianproduct}
The \emph{$d$-dimensional Cartesian product} of graphs $G_1, G_2, ..., G_d$, denoted $G_1\times G_2 \times \cdots \times G_d$, is the graph with vertex set $V(G_1)\times V(G_2) \times \cdots \times V(G_d)$, in which vertices $(u_1, ..., u_i, ..., u_d)$ and $(v_1, ..., v_i, ..., v_d)$ are adjacent if and only if $(u_i,v_i)\in E(G_i)$ and $u_j=v_j$ for all $j \neq i$.
\end{definition}

The graphs $G_1, G_2, ..., G_d$ are called the \emph{factors} of $G_1\times G_2 \times \cdots \times G_d$.
Observe that $G_1\times G_2 \times \cdots \times G_d$ contains $\prod_{j \neq i} |V(G_j)|$ disjoint copies of $G_i$, which form dimension $i$.
We define now some of the basic factor graphs that will be considered.

\begin{definition}
\label{def:path}
The \emph{path} of $k$ vertices, denoted $P_k$, is a graph such that $V(P_k)=\{0, 1, \ldots, k-1\}$ and where $E(P_k)=\{(i, i+1) :  i\in [0,k-2]\}$ .
\end{definition}

\begin{definition}
\label{def:complete}
The \emph{complete graph} (a.k.a. the clique) of $k$ vertices, denoted $K_k$, is a graph such that $V(K_k)=\{0, 1, \ldots, k-1\}$ and where $E(K_k)=\{(i,j): (j\neq i) \wedge (i, j\in V(K_k))\}$.
\end{definition}

\begin{definition}
\label{def:Ncomplete}
The \emph{r-complete graph} of $k$ vertices denoted $rK_k$, is a graph such that $V(rK_k)=\{0, 1, \ldots, k-1\}$ and where $E(rK_k)$ is a multiset such that each pair of vertices $i,j\in V(rK_k)$ is connected with $r$ parallel edges. (i.e., each $e\in E(rK_k)$ has multiplicity $r$).
\end{definition}

Using these and other graphs as factors, we will define, across the text, different $d$-dimensional Cartesian product graphs. For convenience, for these graphs we will use the general notation $G^{(d)}_{k_1, \ldots, k_d}$, where $G$ is the name of the graph, the superscript $(d)$ means that it is a $d$-dimensional graph, and $k_1,   \ldots k_d$ are the number of vertices in each dimension.
(Superscript and subscripts may be omitted when clear from the context.)
It will always hold that $k_1 \geq k_2 \geq \ldots \geq k_d$, i.e., the factor graphs are sorted by decreasing number of vertices.
We will often use $n$ to denote the number of nodes a the graph $G^{(d)}_{k_1, \ldots, k_d}$, i.e., $n=k_1k_2\cdots k_d$, and
we will always use $\alpha$ to denote the index of the lowest dimension with an even number of vertices (if there is no such dimension, $\alpha=d$, where $d$ is the index of the lowest dimension).
According to this notation we will present different $d$-dimensional product graphs as follows.

\begin{definition}
\label{def:array}
    The \emph{d-dimensional array}, denoted $A^{(d)}_{k_1, \ldots, k_d}$, is the Cartesian product of $d$ paths of $k_1,   \ldots, k_d$ vertices, respectively.
    I.e., $A^{(d)}_{k_1, \ldots, k_d} = P_{k_1} \times P_{k_2} \times \cdots \times P_{k_d}$.
\end{definition}

\begin{definition}
\label{def:rhamming}
    The \emph{d-dimensional $r$-Hamming graph}, denoted  $rH^{(d)}_{k_1, \ldots, k_d}$, is the Cartesian product of $d$ $r$-complete graphs of $k_1,   \ldots, k_d$ nodes, respectively.
    I.e., $rH^{(d)}_{k_1, \ldots, k_d} = rK_{k_1}\times rK_{k_2} \times \cdots \times rK_{k_d}$.
\end{definition}

Observe that the \emph{Hamming graph} \cite{extremalsets} is the particular case of the $r$-Hamming graph, with $r=1$. For brevity, we use $H^{(d)}_{k_1, \ldots, k_d}$ instead of $1H^{(d)}_{k_1, \ldots, k_d}$, to denote the Hamming graph.

\subsection{Boundaries and partitions}

We define now the dimension-normalized boundary \cite{azizo}.

\begin{definition}
Let $G^{(d)}_{k_1, \ldots, k_d}$ be a $d$-dimensional product graph and $S(G)$ a subset of $V(G)$. Then, the \emph{dimension-normalized boundary} of
$S(G)$, denoted $B_G(S)$, is defined as
\begin{equation}
\label{eq:BX}
    B_G(S) = \frac{|\partial^G_1 S|}{\sigma_1} + \frac{|\partial^G_2 S|}{\sigma_2} + \ldots + \frac{|\partial^G_d S|}{\sigma_d},
\end{equation}
where, for each $i\in [1,d]$, $\partial^G_i$ is $\partial^G$ applied to the dimension $i$ of $G$ and
\begin{equation}\label{eq:cs}\sigma_i = \Bigg\{\begin{array}{ll}
        k_i^2   & \text{if $k_i$ is even} \\
        k_i^2-1 & \text{if $k_i$ is odd.} \\
\end{array} \end{equation}
\end{definition}

\begin{observation}
\label{o-dims}
For $rH^{(d)}_{k_1, \ldots, k_d}$, any subset $S$ of nodes, and any dimension $i$, it holds that $|\partial^{rH}_i S| = r \cdot |\partial^H_i S|$. Hence,

    \begin{eqnarray*}
     \label{eq:BXcorol}
        B_{rH}(S) &=& \frac{|\partial^{rH}_1 S|}{\sigma_1} + \cdots + \frac{|\partial^{rH}_d S|}{\sigma_d} \\
            &=& r \left(\frac{|\partial^{H}_1 S|}{\sigma_1} + \cdots + \frac{|\partial^{H}_d S|}{\sigma_d} \right) \\
            &=& r B_H(S).
    \end{eqnarray*}
\end{observation}
Let us define the lexicographic-order. Consider graph $H^{(d)}_{k_1, \ldots, k_d}$, we say that vertex $x=(x_1,x_2, \ldots, x_d)$ precedes vertex $y=(y_1,y_2, \ldots, y_d)$ in \emph{lexicographic-order} if there exists an index $i \in [1,d]$ such that $x_{i} < y_{i}$ and $x_j=y_j$ for all $j < i$. Azizoglu and Egecioglu \cite{extremalsets} proved the following result.

\begin{theorem}[\cite{extremalsets}]
\label{theo:bx}
Consider a d-dimensional Hamming graph $H^{(d)}_{k_1, \ldots, k_d}$, with $k_1 \geq k_2 \geq \cdots \geq k_d$. Let $S$ be any subset of $V(H)$ and $\bar{S}$ the set of first $|S|$ vertices of $H$ in lexicographic-order\footnote{Observe that we have reversed the ordering of dimensions with respect to the original theorem from Azizoglu and Egecioglu.},
    then $B_H(\bar{S})\leq B_H(S)$.
\end{theorem}

Table \ref{tab:notation} summarizes the basic notation used in this paper.

\begin{table}[bt]
    \caption{Basic Notation}
    \label{tab:notation}
    \centering
        \begin{tabular}{|c|l|}
                \hline
                    \multicolumn{2}{|c|}{\textbf{Notation}}\\
                \hline
                    $(B)BW(G)$ & Bisection (Band)Witdh of graph $G$ \\
                \hline
                    $n$ & Number of nodes in graph $G$ \\
                \hline
                    $k_i$ &  Number of nodes in dimension $i$ \\
                \hline
                    $d$ & Dimension index \\
                \hline
                    $G^{(d)}_{k_1, \ldots, k_d}$ & Graph $G$ with $d$ dimensions of sizes $k_1,k_2,\ldots, k_d$ \\
                \hline
                    $\alpha$ & Lowest index of an even $k_i$ \\
                \hline
                    $\Uppsi(\alpha)$ & Bisection Width of a $d$-dimensional array \\
                \hline
                    $\partial^gS(G)$ & Edges connecting $S(G)$ and $V(G) \setminus S(G)$ \\
                \hline
                    $B_g(S)$ & Dimension normalized boundary of $S(G)$ \\
                \hline
                    $CC(G)$ & Central Cut of graph $G$\\
                \hline
                    $\beta_r(G)$ & Normalized congestion of graph $G$ of multiplicity $r$ \\
                \hline
                    $m_r(G)$ & Congestion of graph $G$ with multiplicity $r$ \\
                \hline
                    $\cal{E}$ & Set of all possible embeddings $M_r$ of $rK_n$ onto $G$ \\
                \hline
                    $T$ & Links capacity \\
                \hline
                    $s$ & Switching capacity \\
                \hline

        \end{tabular}
\end{table}

\section{Bounds on the bisection width of product graphs}
\label{sec:bounds}

In this section we present general bounds on the bisection width of product graphs as well as presenting two important parameters, the normalized congestion and the central cut, which are used to obtain them. These bounds will be used in the upcoming sections to find
the bisection width of several instances of product graphs.

\subsection{Lower bound}

We start by defining the normalized congestion of a graph.
Let $G$ be a graph with $n$ nodes. Then, an \emph{embedding} of graph $rK_n$ onto $G$ is a mapping of the edges of $rK_n$ into paths in $G$.
We define the \emph{congestion of $G$ with multiplicity $r$}, denoted $m_r(G)$, as the minimum (over all such embeddings) of the maximum number of embedded paths that contain an edge from $G$. To formally define this concept, we first define the congestion of an edge $e\in E(G)$ under the
embedding $M_r$ of $rK_n$ onto $G$, denoted $c_{M_r}(e)$, as
\begin{equation}
\label{eq:congestion_mi}
    c_{M_r}(e) = |\{e'\in E(rK_n):e\in M_r(e')\}|.
\end{equation}
(Observe that $M_r(e') \subseteq E(G)$ is a path in $G$.) Then, the congestion $m_r(G)$ is
    \begin{equation}
    \label{eq:congestion_arr}
        m_r(G) = \min_{M_r\in {\cal E}} \max_{e\in E(G)} \{c_{M_r}(e)\},
    \end{equation}
where ${\cal E}$ is the set of all possible embeddings of $rK_n$ onto $G$.
Then, using Eqs. (\ref{eq:congestion_arr}) and (\ref{eq:cs}), we define the normalized congestion with multiplicity $r$ of $G$ as
\begin{equation}
    \label{eq:norm_congestion}
    \beta_r(G)= \frac{m_r(G)}{\sigma_n}.
\end{equation}

Having defined the normalized congestion, we proceed to extend Theorem \ref{theo:bx} to $r$-Hamming graphs.
\begin{theorem}
\label{theo:rbx}
Consider a d-dimensional $r$-Hamming graph $rH^{(d)}$. Let $S$ be any vertex subset of $V(rH^{(d)})$ and $\bar{S}$ the set of first $|S|$ vertices of $rH^{(d)}$ in lexicographic order, then $B_{rH}(\bar{S})\leq B_{rH}(S)$.
\end{theorem}

\begin{proof}
We prove the theorem by contradiction.
Assume that there is a set of vertices $X \neq \bar{S}$ such that $|X| = |\bar{S}|$ and $B_{rH}(\bar{S}) > B_{rH}(X)$. Then,
applying Observation~\ref{o-dims} to both $X$ and
$\bar{S}$, we obtain that
  \begin{equation}
    \label{eq:BXcorol2}
        B_H(\bar{S})  = \frac{B_{rH}(\bar{S})}{r} >  \frac{B_{rH}(X)}{r} = B_H(X),
    \end{equation}
which contradicts Theorem~\ref{theo:bx} and proves the theorem.
\end{proof}

We now present the following lemma.

\begin{lemma}
\label{lem:shortbrh}
    Let  $\bar{S}$ be a subset of the vertices of graph $rH^{(d)}_{k_1,k_2,\ldots, k_d}$, such that $\bar{S}$
    are the first $\lfloor \frac{n}{2} \rfloor$ vertices of $rH$ in lexicographic order, and $n$ is the number of vertices of $rH$.
    Then, the dimension-normalized boundary of $\bar{S}$ is
    $$B_{rH}(\bar{S})=\frac{r}{4}\Uppsi(\alpha).$$
\end{lemma}

\begin{proof}
We will derive first the value of $B_{H}(\bar{S})$, and then use Observation~\ref{o-dims} to prove the claim.
It was shown in \cite{azizo}, that $\partial^H_i \bar{S} = \emptyset$ for all $i > \alpha$.\footnote{Observe that they use reverse lexicographic order and sort dimensions in the opposite order we do.}
The number of edges in each dimension $i\in\ [1, \alpha]$ on the boundary of $\bar{S}$ in $H$ is
    \begin{equation}
    \label{eq:partials}
        |\partial^{H}_i \bar{S}| = \Bigg\{
            \begin{array}{ll}
                    \frac{k_i}{2}(\prod^d_{j=i+1}{k_j})\frac{k_i}{2}  & \text{if $k_i$ is even} \\
                    \frac{k_i-1}{2}(\prod^d_{j=i+1}{k_j})\frac{k_i+1}{2} & \text{if $k_i$ is odd.} \\
            \end{array}
    \end{equation}
Then, from the definition of $B_{H}(\bar{S})$, we obtain that
\begin{eqnarray*}
\label{eq:BXlem2}
            B_{H}(\bar{S})  & = & \frac{\frac{k_1-1}{2}(\prod^d_{j=2}{k_j})\frac{k_1+1}{2}}{k_1^2-1} + \frac{\frac{k_2-1}{2}(\prod^d_{j=3}{k_j})\frac{k_2+1}{2}}{k_2^2-1}\\
            && + \cdots + \frac{\frac{k_\alpha}{2}(\prod^d_{j=\alpha+1}{k_j})\frac{k_\alpha}{2}}{k_\alpha^2}\\
            & = & \frac{\prod^d_{j=2}{k_j}}{4} + \frac{\prod^d_{j=3}{k_j}}{4} + \cdots + \frac{\prod^d_{j={\alpha+1}}{k_j}}{4}\\
            & = & \frac{\sum_{i=1}^\alpha{C_i}}{4}=\frac{\Uppsi(\alpha)}{4}.
\end{eqnarray*}

Finally, from Observation~\ref{o-dims}, we derive $$B_{rH}(\bar{S}) = r B_{H}(\bar{S}) = \frac{r}{4}\Uppsi(\alpha).$$
\end{proof}

 Using Definition~\ref{def:cartesianproduct}, Lemma~\ref{lem:shortbrh}, and Eq.~(\ref{eq:congestion_arr}), we obtain the following theorem.

\begin{theorem}
\label{theo:bisdef}
    Let $G=G_1\times\ldots\times G_d$, where $|V(G_i)|=k_i$ and $k_1\geq k_2\geq \ldots \geq k_d$. Let $\beta_r(G_i)$ be the normalized congestion with multiplicity $r$ of $G_i$  (for any $r$),  for all $i\in [1, d]$. Consider any  subset $S \subset V(G)$ and the subset $\bar{S}$ which contains the first $|S|$ vertices of $G$, in lexicographic order. Then,
    $$B_{rH}(\bar{S})\leq \sum^d_{i=1}{\beta_r(G_i)|\partial^G_i S|}$$
\end{theorem}

\begin{proof}
    First, observe that, for any $S_i \subset V(G_i)$,
    \begin{equation}
    \label{eq:bisdef_1}
        |\partial^{rK_{k_i}} S_i| \leq m_r(G_i)\cdot |\partial^{G_i} S_i|.
    \end{equation}
    Then, for $S \subset V(G)$ as defined,
    $$
     |\partial^{rH}_i S| \leq m_r(G_i)\cdot |\partial^{G}_i S|.
    $$
    Finally, using Theorem~\ref{theo:rbx}, we can state that
    \begin{eqnarray*}
    B_{rH}(\bar{S}) & \leq & B_{rH}(S) \\
    & \leq & m_r(G_1)\frac{|\partial^G_1 S|}{\sigma_1} +\cdots+ m_r(G_d)\frac{|\partial^G_d S|}{\sigma_d}\\
    & = & \beta_r(G_1)|\partial^G_1 S| +\cdots+ \beta_r(G_i)|\partial^G_d S|.
\end{eqnarray*}
%
%
%
%
%
\end{proof}

From this theorem, we derive a corollary for the case of $|S|=\lfloor\frac{n}{2} \rfloor$:

\begin{corollary}
\label{cor:bisdef}
    Let $G=G_1\times\ldots\times G_d$, where $|V(G_i)|=k_i$ and $k_1\geq k_2\geq \ldots \geq k_d$. Let $\beta_r(G_i)$ be the normalized congestion with multiplicity $r$ of $G_i$ (for any $r$), for $i\in [1, d]$. Consider any subset $S \subset V(G)$ such that $|S|=\lfloor\frac{|V(G)|}{2} \rfloor$. Then
    $$\frac{r}{4}\Uppsi(\alpha) \leq \sum^d_{i=1}{\beta_r(G_i)|\partial^G_i S|}.$$
\end{corollary}

\begin{corollary}
\label{cor:bisexact}
    Let $G=G_1\times\ldots\times G_d$, where $|V(G_i)|=k_i$ and $k_1\geq k_2\geq \ldots \geq k_d$. Let $\beta_r(G_i)=\beta$ be the normalized congestion with multiplicity $r$ of $G_i$ (for any $r$), for $i\in [1, d]$. Consider any subset $S \subset V(G)$ such that $|S|=\lfloor\frac{|V(G)|}{2} \rfloor$. Then
    $$\frac{r}{4\beta}\Uppsi(\alpha) \leq BW(G).$$
\end{corollary}

\subsection{Upper bound}

Having proved the lower bound on the bisection width, we follow with the upper bound. We define first the central cut of a graph $G$.

Consider a graph $G$ with $n$ nodes, and a partition of $V(G)$ into three sets $S^-$, $S^+$, and $S$, such that $|S^-|=|S^+|=\lfloor\frac{n}{2}\rfloor$ (observe that if $n$ is even then
$S=\emptyset$, otherwise $|S|=1$). Then, the \emph{central cut of $G$}, denoted $CC(G)$, is
$$
\min_{\{S^-, S^+, S\}} \max \{|\partial^G S^-|,|\partial^G S^+|\}.
$$
Observe that, for even $n$, the central cut is the bisection width.
Now we use the definition of central cut in the following theorem.

\begin{theorem}
\label{theo:ubound}
Let $G=G_1\times\ldots\times G_d$, where $|V(G_i)|=k_i$ and $k_1\geq k_2\geq \ldots \geq k_d$.
Then,
$$ BW(G) \leq \max_i{\{CC(G_i)\}} \cdot \Uppsi(\alpha). $$
\end{theorem}

\begin{proof}
\label{proof:ubound}
It was shown in \cite{azizo} how to bisect $A^{(d)}$ by cutting exactly $BW(A^{(d)})=\Uppsi(\alpha)$ links.
Furthermore, this bisection satisfies that, if the paths $P_{k_i}$ in dimension $i$ are cut, each of them can be partitioned into subpaths
$P^+$ and $P^-$ of size $\lfloor\frac{k_i}{2}\rfloor$ (connected by a link if $k_i$ is even or by a node with links to both if $k_i$ is odd)
so that the cut separates $P^+$ or $P^-$ from the rest of the path. Each path is then cut by removing one link.
We map the sets $S^+$ and $S^+$ of the partition that gives the central cut of $G_i$ to $P^+$ and $P^-$, respectively. Then,
any cut of a paths $P_{k_i}$ in dimension $i$ becomes a cut of $G_i$ with at most $CC(G_i)$ links removed.

Then, if $S$ is the subset of $V(G)$ that ends at one side of the bisection described above, we have that
    \begin{equation}
    \label{eq:ubound_1}
        \frac{|\partial^{G}_i S|}{CC(G_i)} \leq |\partial^{A^{(d)}}_i S|,
    \end{equation}
which also holds if the paths in dimension $i$ are not cut. Applying this to all dimensions, we obtain
    \begin{equation}
    \label{eq:ubound_2}
           \frac{|\partial^{G}_1 S|}{CC(G_1)} +\cdots+ \frac{|\partial^{G}_d S|}{CC(G_d)} \leq BW(A^{(d)})= \Uppsi(\alpha).
    \end{equation}
This yields,
$$ BW(G) \leq |\partial^{G}_1 S| + \cdots+ |\partial^{G}_d S|\leq \max_i{\{CC(G_i)\}} \cdot \Uppsi(\alpha), $$
    proving Theorem \ref{theo:ubound}.
\end{proof}

\section{Bisection width of products of CBTs and paths}
\label{sec:pathtree}

In this section we will obtain the bisection bandwidth of product graphs which result from the Cartesian product of paths and CBTs. We will present, first, the different factor graphs we are using and the product graphs we are bisecting, then, we will compute the congestion and central cut of these factor graphs and, finally, calculate the bisection width of these product graphs.

\subsection{Factor and product graphs}

In this section we will work with paths, which were defined in Section \ref{sec:defs}, and CBTs, which we define now.

\begin{definition}
\label{def:cbt}
     The \emph{complete binary tree} of $k$ vertices, denoted $\mathit{CBT_k}$, is a graph such that $V(\mathit{CBT_k})=\{1, 2, \ldots, k\}$, with $k=2^j-1$ ($j$ is the number of levels of the tree), and where $E(\mathit{CBT_k})=\{(i,j): ((j=2i) \vee (j=2i+1)) \wedge (i\in [1,2^{j-1}-1])\}$.
\end{definition}

Combining these factor graphs through the Cartesian product, we obtain the product networks that we define below.

\begin{definition}
\label{def:pathcbt}
    A \emph{$d$-dimensional mesh-connected trees and paths}, denoted  $\mathit{MCTP}^{(d)}_{k_1,k_2,\ldots, k_d}$, is the Cartesian product of $d$ graphs of $k_1, k_2, \ldots, k_d$ vertices, respectively, where each factor graph is a complete binary tree or a path.
    I.e., $\mathit{MCTP}^{(d)}_{k_1,k_2,\ldots, k_d} = G_{k_1} \times G_{k_2} \times \cdots \times G_{k_d}$, where either $G_{k_i}=\mathit{CBT_{k_i}}$ or $G_{k_i}=P_{k_i}$.
\end{definition}

We also define the \emph{d-dimensional mesh-connected trees} \cite{efernan}, denoted  $\mathit{MCT}^{(d)}_{k_1,k_2,\ldots, k_d}$ as the graph $\mathit{MCTP}^{(d)}_{k_1,k_2,\ldots, k_d}$ in which all the factor graphs are complete binary trees. (Observe that the array is also the special case of $\mathit{MCTP}^{(d)}_{k_1,k_2,\ldots, k_d}$ in which all the factor graphs are paths.)

\subsection{Congestion and central cut of paths and CBTs}

The bisection widths of the aforementioned product graphs can be calculated using the bounds defined in Section \ref{sec:bounds}. To do so, we need to compute first the values of the normalized congestion and central cut of their factor graphs, it is, of a path and of a CBT.

We will start by computing the congestion of a path and of a CBT and, then, their central cuts. We present the following lemma.

\begin{lemma}
\label{lem:congarray}
    The congestion of $P_k$ with multiplicity $r$, denoted $m_r(P_k)$, has two possible values, depending on whether the number of vertices $k$ is even or odd, as follows,
    \begin{equation}\label{eq:congpath}m_r(P_k) = \Bigg\{\begin{array}{ll}
            r\frac{k^2}{4}   & \text{if $k$ is even} \\
            r\frac{k^2-1}{4} & \text{if $k$ is odd} \\
    \end{array} \end{equation}
\end{lemma}

\begin{figure}[!tb]
      \centering
      \subfigure[The $4$-vertex path and clique]{\label{fig:array4}\includegraphics[height=1.7cm]{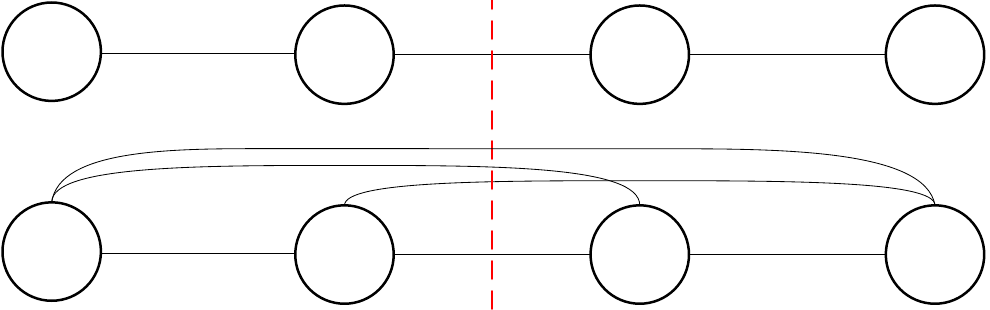}}
      \hfill
      \subfigure[The $5$-vertex path and clique]{\label{fig:array5}\includegraphics[height=1.7cm]{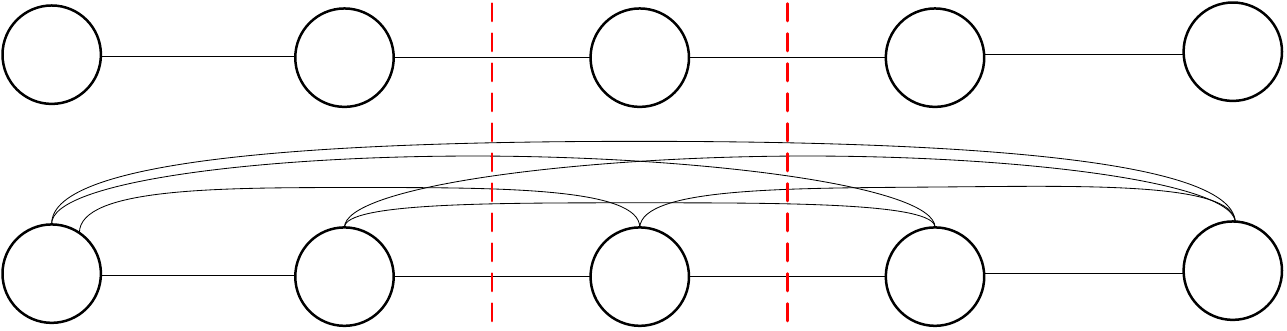}}
      \caption{Paths and possible cuts}
      \label{fig:congarrays}
    \end{figure}

\begin{proof}
This proof is illustrated in Figure~\ref{fig:congarrays} where it can be seen that there are two possible cases, depending on whether $k$ is even or odd.
The congestion $m_r(P_k)$ is defined as the minimum congestion over all embeddings of $rK_k$ onto $P_k$. As there is only one possible path between every pair of vertices, the congestion of an edge will always be the same for any embedding $M_r$ of $rK_k$ into $P_k$. Let $M_r$ be an embedding of $rK_k$ onto $P_k$. Then,
    \begin{equation}
    \label{eq:congestion_path}
        m_r(P_k) = \min_{M\in {\cal E}} \max_{e\in E(P_k)} \{c_{M}(e)\} = \max_{e\in P_k} \{c_{M_r}(e)\}.
    \end{equation}
  If we fix $e=(i,i+1)\in E(P_k)$, $i \in [0,k-1]$, the congestion of $e$ follows the equation:
    \begin{equation}
    \label{eq:congestion_path2}
        c_{M_r}(e) = r(i+1)(k-i-1).
    \end{equation}
    The value of $i$ that maximizes $c_{M_r}(e)$ is $i = \frac{k}{2}-1$. As $k$ is an integer, depending on whether $k$ is even or odd, $\frac{k}{2}$ will be exact or not. Hence, we consider two possible cases,
    \begin{equation}\label{eq:is} i = \Bigg\{\begin{array}{ll}
            \frac{k}{2}-1   & \text{if $k$ is even} \\
            \frac{k-1}{2}-1 & \text{if $k$ is odd.} \\
    \end{array} \end{equation}
    Using these values in Eq.~(\ref{eq:congestion_path2}) leads to the final result

    \begin{equation*}m_r(P_k) = \Bigg\{\begin{array}{ll}
            r\frac{k^2}{4}   & \text{if $k$ is even} \\
            r\frac{k^2-1}{4} & \text{if $k$ is odd} \\
    \end{array} \end{equation*}
\end{proof}

\begin{corollary}
\label{cor:norcongpath}
    The normalized congestion of a path is $\beta_r(P_k) =\frac{r}{4}$.
\end{corollary}

\begin{figure}[!bt]
      \centering
      \includegraphics[height=2.5cm]{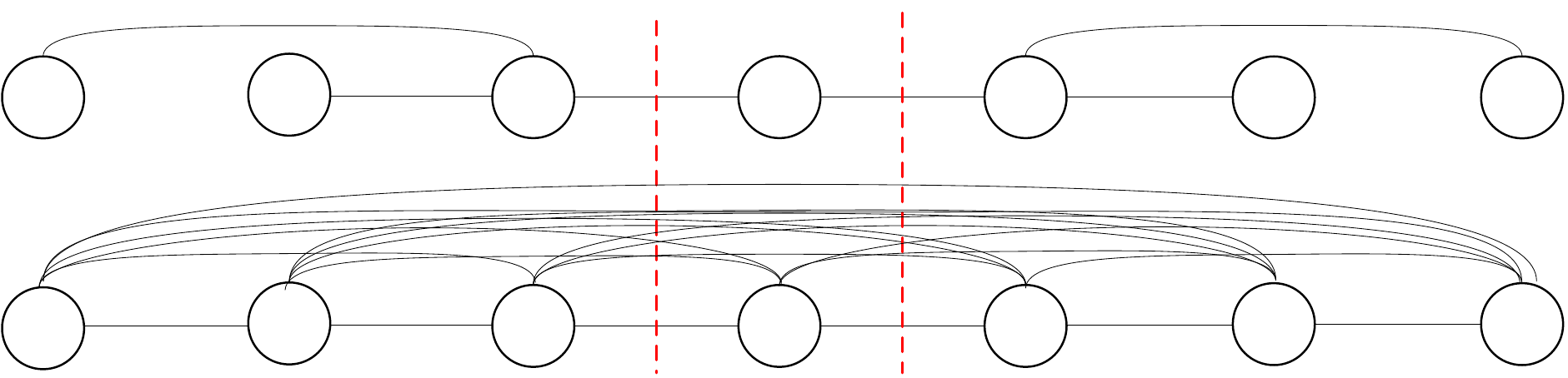}
      \caption{The $7$-vertex complete binary tree and the $7$-vertex clique, with their possible cuts}
      \label{fig:cbt7}
    \end{figure}

The value of the congestion of a CBT will be exactly the same obtained for a path with an odd number of nodes. CBTs share the property of the path of having only one possible routing between two nodes. As can be seen in Figure \ref{fig:cbt7}, the possible cuts are similar. We present Lemma \ref{lem:congcbt} for the congestion of a CBT.

\begin{lemma}
\label{lem:congcbt}
    The congestion of $\mathit{CBT}_k$ with multiplicity $r$, denoted $m_r(\mathit{CBT}_k)$ is
    \begin{equation}\label{eq:congcbt}
        m_r(\mathit{CBT}_k) = r\frac{k^2-1}{4}
    \end{equation}
\end{lemma}

\begin{proof}
    Let $\mathit{CBT}_{2^j-1}$ be a complete binary tree of $j$ levels with $k=2^j-1$ nodes. Whichever edge we cut results on two parts, one of them being another complete binary tree, let us call it $A$ and assume it has $l<j$ levels; and the other being the rest of the previous complete binary tree, let us call it $B$. The number of nodes in $A$ will be $2^l-1$ while the number of nodes in $B$ will be $k-2^l +1$.
    For any embedding $M$ of $rK_k$ into $\mathit{CBT}_k$, the congestion of any edge $e$ follows the equation

    \begin{equation}
    \label{eq:congestion_cbt}
        c_{M_r}(e) = r(2^l -1)(k-2^l +1).
    \end{equation}

    The value of $l$ which maximizes the equation is $l=j-1$, which is equivalent to cut one of the links of the root. This divides the tree into
    subgraphs of sizes $\frac{k+1}{2}$ and $\frac{k-1}{2}$.
%
    Then, the final value for congestion will be
    \begin{equation*}
        m_r(\mathit{CBT}_k) = r\frac{k^2-1}{4}
    \end{equation*}
\end{proof}

\begin{corollary}
\label{cor:norcongCBT}
    The normalized congestion of a CBT is $\beta_r(\mathit{CBT}_k) = \frac{r}{4}$.
\end{corollary}

The central cut of both the path and CBT can be easily deduced from Figures \ref{fig:array4}, \ref{fig:array5} and \ref{fig:cbt7}, being $CC(P_k)=CC(\mathit{CBT}_k)=1$.

\subsection{Bounds on the bisection width of products of CBTs and paths}

Having computed both the congestion and the central cut of the possible factor graphs, we can calculate now the lower and upper bound on the bisection width of a product of CBTs and paths. We will start by the lower bound on the bisection width.

\begin{lemma}
\label{lem:lbdpathcbt}
    The bisection width of a $d$-dimensional mesh-connected trees and paths, $\mathit{MCTP}^{(d)}$, is lower bounded by $\Uppsi(\alpha)$.
\end{lemma}

\begin{proof}
\label{proof:lb_dpathcbt}
    As we can see in Corollaries \ref{cor:norcongpath} and \ref{cor:norcongCBT}, the normalized congestion of both factor graphs is the same value
    $r/4$.
    Then, we can apply Corollary~\ref{cor:bisexact}, so
    \begin{equation}
    \label{eq:dpathcbt_dem3}
        \frac{r}{4r/4}\Uppsi(\alpha)\leq BW(\mathit{MCTP}^{(d)})
    \end{equation}
    which yields,
    \begin{equation}
    \label{eq:dpathcbt_dem4}
        BW(\mathit{MCTP}^{(d)}) \geq \Uppsi(\alpha).
    \end{equation}
\end{proof}

We follow now by presenting an upper bound on the bisection width of $d$-dimensional mesh-connected trees and paths.

\begin{lemma}
\label{lem:ubdpathcbt}
    The bisection width of a $d$-dimensional mesh-connected trees and paths, $\mathit{MCTP}^{(d)}$, is upper bounded by $\Uppsi(\alpha)$.
\end{lemma}

\begin{proof}
\label{proof:ub_dpathcbt}
Obviously, as this graph can also be embedded into a $d$-dimensional array, we can use Theorem \ref{theo:ubound}. We know that the central cut of both CBTs and paths is $1$ independently of their sizes or number of levels, and hence also $\max_i \{ CC(G_{k_i})\}=1$ (where $G_{k_i}$ is either a CBT or a path). Then,
    \begin{equation}
        BW(\mathit{MCTP}^{(d)}) \leq \Uppsi(\alpha).
    \end{equation}
\end{proof}

From the results obtained from Lemma \ref{lem:lbdpathcbt} and Lemma \ref{lem:ubdpathcbt} the proof of Theorem~\ref{theo:bwdpathcbt} follows.

    \begin{theorem}
    \label{theo:bwdpathcbt}
        The bisection width of a $d$-dimensional mesh-connected trees and paths $\mathit{MCTP}^{(d)}_{k_1,k_2,\ldots, k_d}$ is $\Uppsi(\alpha)$.
    \end{theorem}

We can also present the following corollary for the particular case of the $d$-dimensional mesh-connected trees $\mathit{MCT}^{(d)}_{k_1,k_2,\ldots, k_d}$.

\begin{corollary}
\label{cor:bwdcbt}
    The bisection width of the $d$-dimensional mesh-connected trees $\mathit{MCT}^{(d)}_{k_1,k_2,\ldots, k_d}$ is $BW(\mathit{MCT}^{(d)})=\Uppsi(d)$.
\end{corollary}

\section{Products of rings and extended trees}
\label{sec:ringxt}

Similarly to what was done in Section \ref{sec:pathtree}, in this section we will obtain a result for the bisection bandwidth of the product graphs which result from the Cartesian product of rings and extended complete binary trees, a.k.a. XTs.

\subsection{Factor and product graphs}

The factor graphs which are going to be used in this section are rings and XTs. We define them below.

\begin{definition}
\label{def:ring}
The \emph{ring} of $k$ vertices, denoted $R_k$, is a graph such that $V(R_k)=\{0, 1, \ldots, k-1\}$ and where $E(R_k)=\{(i, (i+1)\mod{k}): i\in V(R_k)\}$ .
\end{definition}

\begin{definition}
\label{def:xt}
The \emph{extended complete binary tree} (a.k.a. XT) of $k$ vertices, denoted $X_k$, is a complete binary tree in which the leaves are connected as a path. More formally,
$V(X_k)=V(\mathit{CBT_k})$ and $E(X_k)= E(\mathit{CBT_k}) \cup \{(i,i+1):  i\in [2^{j-1},2^j-2]\}$.
\end{definition}

Combining these graphs as factor graphs in a Cartesian product, we can obtain the three following different kinds of product graphs:

\begin{definition}
\label{def:ringsxt}
    A \emph{$d$-dimensional mesh-connected extended trees and rings}, denoted $\mathit{MCXR}^{(d)}_{k_1,k_2,\ldots, k_d}$, is the Cartesian product of $d$ graphs of $k_1, k_2, \ldots, k_d$ vertices, respectively, where each factor graph is a extended complete binary tree or a ring.
    I.e., $\mathit{MCXR}^{(d)}_{k_1,k_2,\ldots, k_d} = G_{k_1} \times G_{k_2} \times \cdots \times G_{k_d}$, where either $G_{k_i}=X_{k_i}$ or $G_{k_i}=R_{k_i}$.
\end{definition}

\begin{definition}
\label{def:torus}
    The \emph{$d$-dimensional torus}, denoted  $T^{(d)}_{k_1,k_2,\ldots, k_d}$, is the Cartesian product of $d$ rings of $k_1, k_2, \ldots, k_d$ vertices, respectively.
    I.e., $T^{(d)}_{k_1,k_2,\ldots, k_d} = R_{k_1} \times R_{k_2} \times \cdots \times R_{k_d}$.
\end{definition}

And, as happened in Section \ref{sec:pathtree} with $\mathit{MCT}^{(d)}$, we also define the \emph{$d$-dimensional mesh-connected extended trees}, denoted $\mathit{MCX}^{(d)}_{k_1,k_2,\ldots, k_d}$, a special case of $\mathit{MCXR}^{(d)}_{k_1,k_2,\ldots, k_d}$ in which all factor graphs are extended complete binary trees. (The torus is the special case of $\mathit{MCXR}^{(d)}_{k_1,k_2,\ldots, k_d}$ in which all factor graphs are rings.)

\subsection{Congestion and central cut of rings and XTs}

The congestion and central cut of both a ring and an XT are needed to calculate the bounds obtained in Section \ref{sec:bounds}. We present the following lemma for the congestion of a ring.

\begin{lemma}
\label{lem:congring}
The congestion of $R_k$ with multiplicity $r=2$ has two possible upper bounds depending on whether the number of vertices $k$ is even or odd, as follows,
    \begin{equation}\label{eq:congring}m_2(R_k) \leq \Bigg\{\begin{array}{ll}
            1\frac{k^2}{4}   & \text{if $k$ is even} \\
            1\frac{k^2-1}{4} & \text{if $k$ is odd}
    \end{array} \end{equation}
\end{lemma}

\begin{proof}
    While a path had only one possible routing, for $R_k$ we have two possible routes connecting each pair of nodes.
    If we embed $rK_k$, for $r=2$, into $R_k$, we can route each of the parallel edges connecting two nodes through each of the possible routings. This yields,
    \begin{equation*}m_2(R_k)\leq \Bigg\{\begin{array}{ll}
        2\frac{\frac{k}{2}\frac{k}{2}}{2} = \frac{k^2}{4}   & \text{if $k$ is even} \\
        2\frac{\frac{k-1}{2}\frac{k+1}{2}}{2} = \frac{k^2-1}{4}  & \text{if $k$ is odd.}
        \end{array}
    \end{equation*}
\end{proof}

\begin{corollary}
\label{cor:norcongring}
    The normalized congestion with multiplicity $r=2$ of a ring is $\beta_2(R_k) =1/4$.
\end{corollary}

    \begin{figure}[!bt]
      \centering
      \subfigure[The $4$-vertex ring and clique]{\label{fig:torus4}\includegraphics[height=2.2cm]{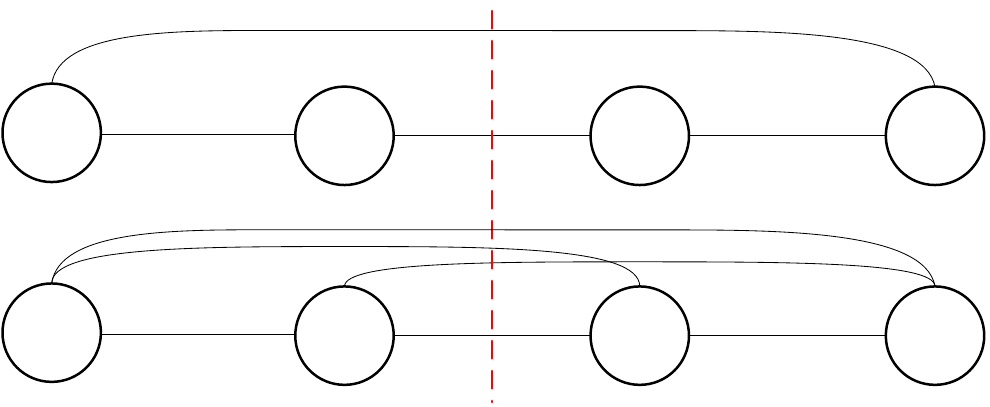}}
      \hfill
      \subfigure[The $5$-vertex ring and clique]{\label{fig:torus5}\includegraphics[height=2.2cm]{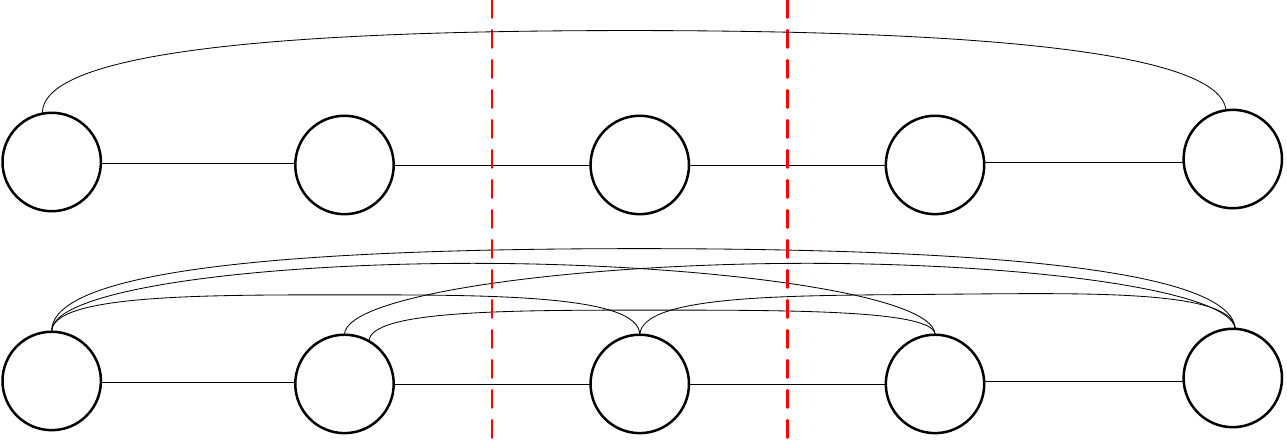}}
      \caption{Rings and possible cuts}
      \label{fig:congtorus}
    \end{figure}

Similarly to what happened with paths and CBTs, the congestion of rings and XTs is the same. The extended complete binary tree $X_k$ has a Hamiltonian cycle \cite{efernan}, so we can find a ring $R_k$ contained onto it. Consequently, the congestion of an XT and a ring with the same number of nodes will be the same. Then, the normalized congestion of both factor graphs will also be the same.

\begin{corollary}
\label{cor:norcongxt}
    The normalized congestion with multiplicity $r=2$ of an XT is $\beta_2(X_k) =1/4$.
\end{corollary}

\begin{figure}[!bt]
      \centering
      \includegraphics[height=3.5cm]{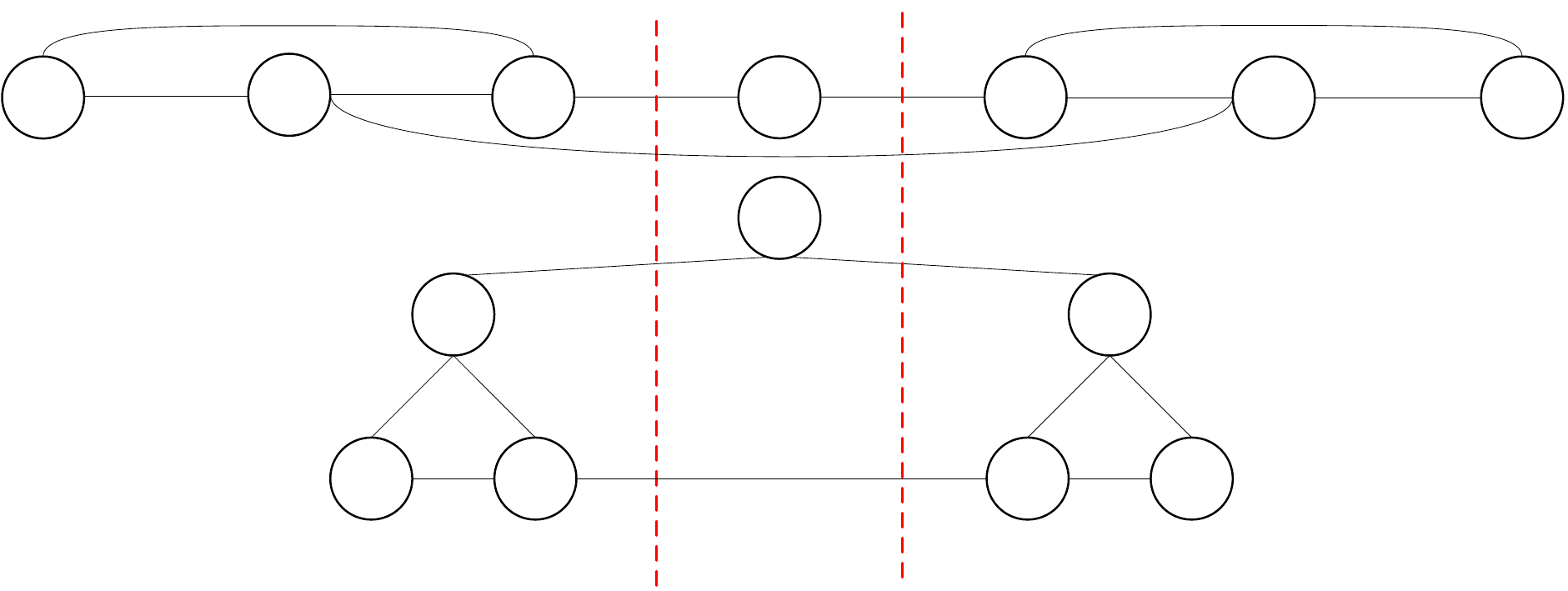}
      \caption{Central cut on a extended complete binary tree}
      \label{fig:ccmcxt}
    \end{figure}

Due to these similarities, central cuts of both graphs are also going to be the same. As can be easily deduced from Figures \ref{fig:torus4}, \ref{fig:torus5} and \ref{fig:ccmcxt}, $CC(R_k)=CC(X_k)=2$.

\subsection{Bounds on the bisection width of products of XTs and rings}

As we did in Section \ref{sec:pathtree}, once we have computed the results for the normalized congestion and central cut of the different factor graphs, we can calculate the lower and upper bounds on the bisection width of products of XTs and rings. We will start by the lower bound on the bisection width presenting the following lemma.

\begin{lemma}
\label{lem:lbdxtring}
    The bisection width of a $d$-dimensional mesh-connected XTs and rings, $\mathit{MCXR}^{(d)}$, is lower bounded by $2\Uppsi(\alpha)$.
\end{lemma}

\begin{proof}
\label{proof:lb_dxtring}
The normalized congestion of both factor graphs is $\beta_2(R_k)= \beta_2(X_k) =\frac{1}{4}$.
Then, applying Corollary \ref{cor:bisexact} with $r=2$,
    \begin{equation}
    \label{eq:dxtring_dem}
        \frac{2}{4(1/4)}\Uppsi(\alpha) \leq BW(\mathit{MCXR}^{(d)})
    \end{equation}
Which yields,
    \begin{equation}
    \label{eq:dxtring_dem4}
        BW(\mathit{MCXR}^{(d)}) \geq 2\Uppsi(\alpha).
    \end{equation}
\end{proof}

We calculate now the upper bound on the bisection width of a $d$-dimensional mesh-connected rings and XTs.

\begin{lemma}
\label{lem:ubdxtring}
    The bisection width of a $d$-dimensional, $\mathit{MCXR}^{(d)}$, is upper bounded by $2\Uppsi(\alpha)$.
\end{lemma}

\begin{proof}
\label{proof:ub_dxtring}
    The $d$-dimensional mesh-connected XTs and rings graph can also be embedded into a $d$-dimensional array, so then, we can use Theorem~\ref{theo:ubound}. As happened with the congestion, the value of the central cut of both XTs and rings is the same, concretely, $CC(R_k)=CC(X_k)=2$, independently of their sizes or number of levels. Hence, $\max_i \{ CC(G_{k_i})\}=2$ (where $G_{k_i}$ is either a ring or an XT). Then,
    \begin{equation}
        BW(\mathit{MCXR}^{(d)}) \leq  2\Uppsi(\alpha).
    \end{equation}
\end{proof}

    From Lemma \ref{lem:lbdxtring} and Lemma \ref{lem:ubdxtring}, Theorem~\ref{theo:bwdxtring} follows.

\begin{theorem}
\label{theo:bwdxtring}
    The bisection width of a $d$-dimensional mesh-connected XTs and rings $\mathit{MCXR}^{(d)}_{k_1,k_2,\ldots, k_d}$ is $2\Uppsi(\alpha)$.
\end{theorem}

   From the bisection width of the $d$-dimensional mesh-connected XTs and rings, we can derive the following corollaries for the particular cases where all the factor graphs are rings, Torus $T^{(d)}$, or XTs, mesh-connected extended trees $\mathit{MCX}^{(d)}$.

\begin{corollary}
\label{cor:bwdtorus}
    The bisection width of the $d$-dimensional torus $T^{(d)}_{k_1,k_2,\ldots, k_d}$ is $BW(T^{(d)})=2\Uppsi(\alpha)$.
\end{corollary}

\begin{corollary}
\label{cor:bwdcbt}
    The bisection width of the $d$-dimensional mesh-connected extended trees $\mathit{MCX}^{(d)}_{k_1,k_2,\ldots, k_d}$ is $BW(\mathit{MCX}^{(d)})=2\Uppsi(d)$.
\end{corollary}

\section{BCube}
\label{sec:bcube}

We devote this section to obtain bounds on the bisection width of a $d$-dimensional BCube\cite{bcube}. BCube is different from the topologies considered in the previous sections because it is obtained as the combination of basic networks formed by a collection of $k$ nodes (servers) connected by a switch.
These factor networks are combined into multidimensional networks in the same way product graphs are obtained from their factor graphs. This allows us to study the BCube as an special instance of a product network. The $d$-dimensional BCube can be obtained as the $d$ dimensional product of
one-dimensional BCube networks, each one of $k$ nodes.

\subsection{Factor and product graphs}

We first define a \emph{Switched Star network} and how a $d$-dimensional BCube network is built from it.

\begin{definition}
\label{def:bcube0}
A \emph{Switched Star network} of $k$ nodes, denoted $\mathit{SS}_{k}$, is composed of $k$ nodes connected to a $k$-ports switch. It can be seen as a complete graph $K_k$ where all the edges have been replaced by a switch.
\end{definition}

Combining this network $d$ times as a factor network in the Cartesian product, we obtain a $d$-dimensional BCube.

\begin{definition}
\label{def:bcubed}
A \emph{$d$-dimensional BCube}, denoted by $\mathit{BC}^{(d)}_{k}$, is the Cartesian product of $d$ $\mathit{SS}_{k}$ (the switches are not considered nodes for the Cartesian product).
    I.e., $\mathit{BC}^{(d)}_{k} = \mathit{SS}_{k}\times \mathit{SS}_{k} \times \cdots \times \mathit{SS}_{k}$.

\end{definition}

$\mathit{BC}^{(d)}_{k}$ can also be seen as a $d$-dimensional homogeneous array where all the edges in each path have been removed and replaced by a switch where two nodes $(u_1, ..., u_i, ..., u_d)$ and $(v_1, ..., v_i, ..., v_d)$ are connected to the same switch if and only if $(u_i\neq v_i)$ and $u_j=v_j$ for all $j \neq i$.


The main reason for obtaining the bisection width of a $d$-dimensional BCube is to be able to bound its bisection bandwidth. However, as the $d$-dimensional BCube is not a typical graph, the bisection width can have different forms depending on where the communication bottleneck is located in a BCube network.

We present two possible models for $\mathit{SS}_k$. The first one, Model-A or \emph{star-like model,} denoted by $\mathit{SSA}_{k}$, consists of $k$ nodes connected one-to-one to a virtual node which represents the switch. The second one, Model-B or \emph{hyperlink model,} denoted by $\mathit{SSB}_{k}$, consists of $k$ nodes connected by a hyperlink.
While the two presented models are logically equivalent to a complete graph, they have a different behavior from the traffic point of view. We show this with two simple examples.

Let us consider that we have a $\mathit{SS}_{3}$ where the links have a speed of $100$ Mbps while the switch can switch at $1$ Gbps. Under these conditions, the links become the bottleneck of the network and, even when the switches would be able to provide a bisection bandwidth of $1$ Gbps, the effective bisection bandwidth is only of $200$ Mbps in both directions.

Consider the opposite situation now, where the BCube switch only supports $500$ Mbps of internal traffic while the links transmit at $1$ Gbps. In this case, the switches are the bottleneck of the network and the bisection bandwidth is only $500$ Mbps, although the links would be able to support up to $2$ Gbps.

The first example illustrates an scenario where we would bisect the network by removing the links that connect the servers to the switches, which corresponds to Model $A$. On the other hand, what we find in the second example is a typical scenario for Model B, where we would do better by removing entire switches when bisecting the network. In particular, being $s$ the switching capacity of a switch, and $T$ the traffic supported by a link, we will choose Model-A when $s\geq \lfloor \frac{k}{2}\rfloor \cdot 2T$ and Model-B when $s\leq 2T$. (Note that this does not cover the whole spectrum of possible values of $s$, $T$, and $k$.)

\subsection{Congestion and central cut of BCube}

We will compute now the congestion and central cut of both models in order to be able to calculate the respective lower and upper bounds.
We start by the congestion and central cut of Model-A.

    \begin{figure}[!bt]
      \centering
      \subfigure[Model A: $5$-node Star-like BCube]{\label{fig:modelA}\includegraphics[height=1.4cm]{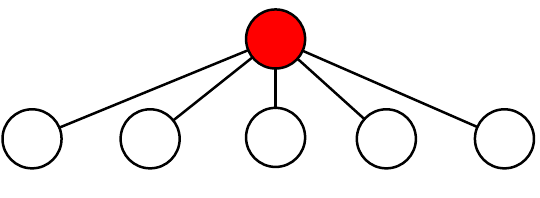}}
      \hfill
      \subfigure[Congestion of a $5$-node Star-like BCube]{\label{fig:modelAcong}\includegraphics[height=1.3cm]{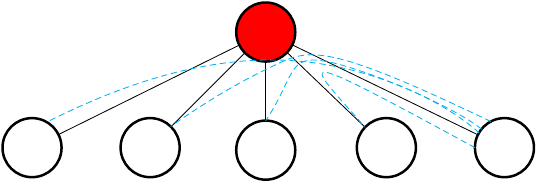}}
      \hfill
      \subfigure[Central cut of a $5$-node Star-like BCube]{\label{fig:modelAcc}\includegraphics[height=1.5cm]{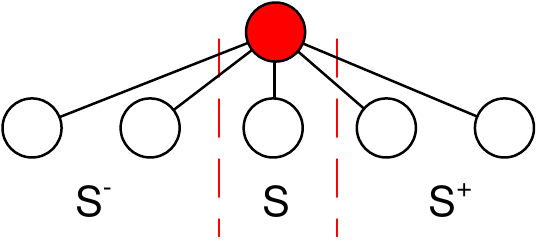}}
      \caption{Model A of a 5-node BCube and its congestion and central cut}
      \label{fig:modela}
    \end{figure}

Model-A is also called star-like model. The name of star-like comes from the fact that the factor graph can be seen as a star with the switch in the center. If we set $r=1$, the congestion of every link of the star is easily found to be $m_r(\mathit{SSA}_{k})=k-1$\footnote{Note that in the computation of the congestion, the switch is not considered a node of the graph.} as shown in Figure \ref{fig:modelAcong}.

\begin{corollary}
\label{cor:norcongxt}
    The normalized congestion of $\mathit{SSA}_{k}$ is
    \begin{equation*}
        \beta_r(\mathit{SSA}_{k}) = \Bigg\{
        \begin{array}{ll}
            \frac{k-1}{k^2} & \text{if $k$ is even} \\
            \frac{1}{k+1}  & \text{if $k$ is odd.} \\
        \end{array}
    \end{equation*}

\end{corollary}

The central cut, which is also trivial and can be found in Figure \ref{fig:modelAcc}, will depend on whether the number of nodes $k$ is even or odd,
    \begin{equation*}
        CC(\mathit{SSA}_{k}) = \Bigg\{
        \begin{array}{ll}
            \frac{k}{2} & \text{if $k$ is even} \\
            \frac{k-1}{2}  & \text{if $k$ is odd.} \\
        \end{array}
    \end{equation*}

Having computed the congestion and the central cut for Model-A, we will compute them now for Model-B. We also call Model-B hyperlink model\footnote{This model is quite similar to the one proposed by Pan in \cite{buses}.} due to the fact that all the servers from the BCube are connected by a hyperlink so no switch is needed.

Calculating the congestion of a Model-B BCube will be easy then. If we set $r=1$ there will be only one edge to be removed, the congestion of the graph will be total amount of edges of its equivalent $K_k$, i. e., $m_r(\mathit{SSB}_{k})=\frac{k(k-1)}{2}$.

\begin{corollary}
\label{cor:norcongxt}
    The normalized congestion of $\mathit{SSB}_{k}$ is
    \begin{equation*}
        \beta_r(\mathit{SSB}_{k}) = \Bigg\{
        \begin{array}{ll}
            \frac{k-1}{2k} & \text{if $k$ is even} \\
            \frac{k}{2(k+1)}  & \text{if $k$ is odd.} \\
        \end{array}
    \end{equation*}

\end{corollary}

\begin{figure}[!bt]
      \centering
      \subfigure[Model B: 5-node hyperlink BCube]{\label{fig:modelB}\includegraphics[height=1.4cm]{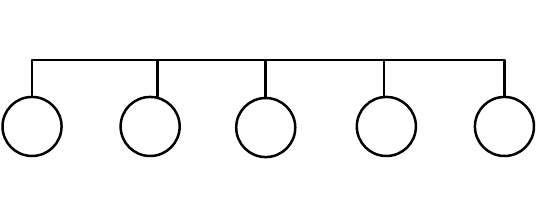}}
      \hfill
      \subfigure[Congestion of a 5-node hyperlink BCube]{\label{fig:modelBcong}\includegraphics[height=1.6cm]{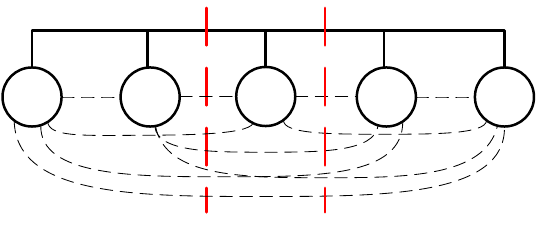}}
      \hfill
      \subfigure[Central cut of a 5-node hyperlink BCube]{\label{fig:modelBcc}\includegraphics[height=1.4cm]{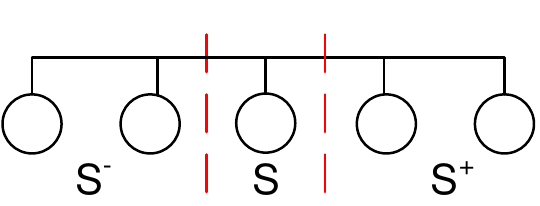}}
      \caption{Model B of a 5-node BCube and its congestion and central cut}
      \label{fig:modelb}
\end{figure}

As for Model-A, the central cut is easily computed. As there is only one hyperlink, its central cut will be $CC(\mathit{BCB}_{k})=1$. Both $m_r(\mathit{SSB}_{k})$ and $CC(\mathit{BCB}_{k})$ are shown in Figures \ref{fig:modelBcong} and \ref{fig:modelBcc}.

\subsection{Bounds on the bisection width of BCube}

Having computed the congestion and central cut of both models, we can calculate the lower and upper bounds on the bisection width of each one of them.

We will start by the lower and upper bounds on the bisection width of model A and, then, we will calculate both bounds for model B.

We first present the following lemma for the lower bound on the bisection width of a Model-A BCube.

\begin{lemma}
\label{lem:lbdbcube}
    The bisection width of a Model-A $d$-dimensional BCube, $\mathit{BCA}^{(d)}_{k}$, is lower bounded by $\frac{k^{d+1}}{4(k-1)}$ if $k$ is even, and by $\frac{k+1}{4}\frac{k^d-1}{k-1}$ if $k$ is odd.
\end{lemma}

\begin{proof}
\label{proof:lbdbcube}
Using the value of the normalized congestion of a Model-A BCube in Corollary \ref{cor:bisexact}, it follows that
$$
    BW(\mathit{BCA}^{(d)}_{k}) \geq \Bigg\{
    \begin{array}{ll}
        \frac{1}{4}\frac{k^2}{k-1}\Uppsi(\alpha) = \frac{k^{d+1}}{4(k-1)}  & \text{if $k$ is even} \\
        \frac{k+1}{4}\Uppsi(\alpha) = \frac{k+1}{4}\frac{k^d-1}{k-1}  & \text{if $k$ is odd} \\
    \end{array}
$$
\end{proof}

After proving the lower bound on the bisection width of a Model-A $d$-dimensional BCube, we follow with the upper bound.

\begin{lemma}
\label{lem:ubdbcube}
    The bisection width of a Model-A $d$-dimensional BCube, $\mathit{BCA}^{(d)}_{k}$, is upper bounded by $\frac{k^d}{2}$ if $k$ is even, and by $\frac{k^d-1}{2}$ if $k$ is odd.
\end{lemma}

\begin{proof}
\label{proof:ubdbcube}
The Cartesian product of Model-A star-like factor graphs can be embedded into a $d$-dimensional array, so Theorem~\ref{theo:ubound} will be extremely useful again. If we use the values of the central cut of Model-A in Theorem~\ref{theo:ubound}, is immediate to compute the following upper bound
$$
        BW(\mathit{BCA}^{(d)}_{k}) \leq \Bigg\{
        \begin{array}{ll}
            \frac{k^d}{2}  & \text{if $k$ is even} \\
            \frac{k^d-1}{2}  & \text{if $k$ is odd.} \\
        \end{array}
$$
\end{proof}

Now, from the combination of Lemma \ref{lem:lbdbcube} and Lemma \ref{lem:ubdbcube} we can state Theorem~\ref{theo:bwdbcube}:

\begin{theorem}
\label{theo:bwdbcube}
The value of the bisection width of a Model-A $d$-dimensional BCube, $\mathit{BCA}^{(d)}_{k}$, is in the interval $[\frac{k^{d+1}}{4(k-1)},\frac{k^d}{2}]$
if $k$ is even, and in the interval $[\frac{k+1}{4}\frac{k^d-1}{k-1},\frac{k^d-1}{2}]$ if $k$ is odd.
\end{theorem}
\begin{corollary}
The bisection bandwidth of a Model-A $d$-dimensional BCube satisfies,
$$
        BBW(\mathit{BCA}^{(d)}_{k}) \in \Bigg\{
        \begin{array}{ll}
            \ [ 2 T \frac{k^{d+1}}{4(k-1)}, 2 T \frac{k^d}{2}]  & \text{if $k$ is even} \\
            \ [ 2 T \frac{k+1}{4}\frac{k^d-1}{k-1}, 2 T \frac{k^d-1}{2}]  & \text{if $k$ is odd.}
        \end{array}
$$
\end{corollary}

Let us calculate now the bounds of a Model-B $d$-dimensional BCube. As we did with Model A, we will first prove the lower bound and then the upper one. For the lower bound we present the following lemma.

\begin{lemma}
\label{lem:lbdbcubeB}
    The bisection width of a Model-B $d$-dimensional BCube, $\mathit{BCB}^{(d)}_{k}$, is lower bounded by $\frac{k^d}{2(k-1)}$ if $k$ is even, and by $\frac{k+1}{2k}\frac{k^d-1}{k-1}$ if $k$ is odd.
\end{lemma}

\begin{proof}
\label{proof:lbdbcubeB}
Like in the case of Model A, we use the value of the normalized congestion of Model B in Corollary \ref{cor:bisexact}. Since all the dimensions have the same size $k$, it follows that

\begin{equation*}
    \label{eq:bcubeB_dem2}
    BW(\mathit{BCB}^{(d)}_{k}) \geq \Bigg\{
    \begin{array}{ll}
        \frac{1}{4}\frac{2k}{k-1}\Uppsi(\alpha) = \frac{k^d}{2(k-1)}  & \text{if $k$ is even} \\
        \frac{1}{4}\frac{2(k+1)}{k}\Uppsi(\alpha) = \frac{k+1}{2k}\frac{k^d-1}{k-1}  & \text{if $k$ is odd} \\
    \end{array}
\end{equation*}

\end{proof}

We present now Lemma \ref{lem:ubdbcubeB} for the upper bound on the bisection width of a Model-B $d$-dimensional BCube.

\begin{lemma}
\label{lem:ubdbcubeB}
    The bisection width of a Model-B $d$-dimensional BCube, $\mathit{BCB}^{(d)}_{k}$, is upper bounded by $\frac{k^d-1}{k-1}$.
\end{lemma}

\begin{proof}
    As for model A, the $d$-dimensional BCube resulting from the Cartesian product of Model-B graphs can be embedded into a $d$-dimensional array. Thanks to this fact, we can use the computed value of its central cut in Theorem~\ref{theo:ubound} to obtain the upper bound on the bisection width,

    \begin{equation*}
    \label{eq:bcubeB_dem}
        BW(\mathit{BCB}^{(d)}_{k}) \leq 1 \cdot \Uppsi(\alpha) = \frac{k^d-1}{k-1}.
    \end{equation*}

\end{proof}

Combining the previous lemmas we can state the following theorem.

\begin{theorem}
\label{theo:bwdbcubeB}
    The value of the bisection width of a Model-B $d$-dimensional BCube, $\mathit{BCB}^{(d)}_{k}$, is in the interval $[\frac{k^d}{2(k-1)},\frac{1-k^d}{1-k}]$
    if $k$ is even, and in the interval $[\frac{k+1}{2k}\frac{k^d-1}{k-1},\frac{k^d-1}{k-1}]$ if $k$ is odd.
\end{theorem}

\begin{corollary}
The bisection bandwidth of a Model-B $d$-dimensional BCube satisfies,
$$
        BBW(\mathit{BCB}^{(d)}_{k}) \in \Bigg\{
        \begin{array}{ll}
            \ [s\frac{k^d}{2(k-1)},s\frac{1-k^d}{1-k}]  & \text{if $k$ is even} \\
            \ [s\frac{k+1}{2k}\frac{k^d-1}{k-1},s\frac{k^d-1}{k-1}]  & \text{if $k$ is odd.}
        \end{array}
$$
\end{corollary}

\section{Conclusions}
\label{s-conclusions}

Exact results for the bisection bandwidth of various $d$-dimensional classical parallel topologies have been provided in this paper. These results consider any number of dimensions and any size, odd or even, for the factor graphs. These multidimensional graphs are based on factor graphs such as paths, rings, complete binary trees or extended complete binary trees. Upper and lower bounds on the bisection width of a $d$-dimensional BCube are also provided.
Some of the product networks studied had factor graphs of the same class, like the $d$-dimensional torus, mesh-connected trees or mesh-connected extended trees, while some other combined different factor graphs, like the mesh connected trees and paths or mesh-connected extended trees and rings.
See Table~\ref{tab:tech} for a summary of the results obtained.

An interesting open problem is how to obtain the exact value of the bisection width of graph obtained by combining paths and rings (cylinders) and other combinations not considered in this paper. Similarly, obtaining an exact result for the bisection bandwidth of the $d$-dimensional BCube remains as an open problem.

\bibliographystyle{IEEEtran}
\bibliography{references}

\providecommand{\bysame}{\leavevmode\hbox to3em{\hrulefill}\thinspace}
\providecommand{\MR}{\relax\ifhmode\unskip\space\fi MR }
\providecommand{\MRhref}[2]{%
  \href{http://www.ams.org/mathscinet-getitem?mr=#1}{#2}
}
\providecommand{\href}[2]{#2}
\begin{thebibliography}{1}

\bibitem{extremalsets}
M.~Cemil Azizoglu and Ömer Egecioglu, \emph{Extremal sets minimizing
  dimension-normalized boundary in hamming graphs}.

\bibitem{azizo}
M~Cemil Azizoglu and Ömer Egecioglu, \emph{The bisection width and the
  isoperimetric number of arrays}, Discrete Applied Mathematics \textbf{138}
  (2004), no.~1-2, 3 -- 12, Optimal Discrete Structures and Algorithms.

\bibitem{efe}
Gui-Liang~Feng Kemal~Efe, \emph{A proof for bisection width of grids}, World
  Academy of Science, Engineering and Technology \textbf{27} (2007), no.~31,
  172 -- 177.

\end{thebibliography}


\begin{thebibliography}{10}
\providecommand{\url}[1]{#1}
\csname url@rmstyle\endcsname
\providecommand{\newblock}{\relax}
\providecommand{\bibinfo}[2]{#2}
\providecommand\BIBentrySTDinterwordspacing{\spaceskip=0pt\relax}
\providecommand\BIBentryALTinterwordstretchfactor{4}
\providecommand\BIBentryALTinterwordspacing{\spaceskip=\fontdimen2\font plus
\BIBentryALTinterwordstretchfactor\fontdimen3\font minus
  \fontdimen4\font\relax}
\providecommand\BIBforeignlanguage[2]{{%
\expandafter\ifx\csname l@#1\endcsname\relax
\typeout{** WARNING: IEEEtran.bst: No hyphenation pattern has been}%
\typeout{** loaded for the language `#1'. Using the pattern for}%
\typeout{** the default language instead.}%
\else
\language=\csname l@#1\endcsname
\fi
#2}}

\bibitem{cylinders}
M.~C. Azizo\u{g}lu and {\"O}.~E\u{g}ecio\u{g}lu, ``The isoperimetric number and
  the bisection width of generalized cylinders,'' \emph{Electronic Notes in
  Discrete Mathematics}, vol.~11, pp. 53--62, 2002.

\bibitem{extremalsets}
------, ``Extremal sets minimizing dimension-normalized boundary in hamming
  graphs,'' \emph{SIAM J. Discrete Math.}, vol.~17, no.~2, pp. 219--236, 2003.

\bibitem{azizo}
------, ``The bisection width and the isoperimetric number of arrays,''
  \emph{Discrete Applied Mathematics}, vol. 138, no. 1-2, pp. 3--12, 2004.

\bibitem{dallyTowles}
W.~Dally and B.~Towles, \emph{Principles and Practices of Interconnection
  Networks}.\hskip 1em plus 0.5em minus 0.4em\relax San Francisco, CA, USA:
  Morgan Kaufmann Publishers Inc., 2003.

\bibitem{dally}
W.~J. Dally, ``Performance analysis of k-ary n-cube interconnection networks,''
  \emph{IEEE Trans. Computers}, vol.~39, no.~6, pp. 775--785, 1990.

\bibitem{duato}
J.~Duato, S.~Yalamanchili, and N.~Lionel, \emph{Interconnection Networks: An
  Engineering Approach}.\hskip 1em plus 0.5em minus 0.4em\relax San Francisco,
  CA, USA: Morgan Kaufmann Publishers Inc., 2002.

\bibitem{efe}
K.~Efe and G.-L. Feng, ``A proof for bisection width of grids,'' \emph{World
  Academy of Science, Engineering and Technology}, vol.~27, no.~31, pp. 172 --
  177, 2007.

\bibitem{efernan3}
K.~Efe and A.~Fern{\'a}ndez, ``Products of networks with logarithmic diameter
  and fixed degree,'' \emph{IEEE Trans. Parallel Distrib. Syst.}, vol.~6,
  no.~9, pp. 963--975, 1995.

\bibitem{efernan}
------, ``Mesh-connected trees: A bridge between grids and meshes of trees,''
  \emph{IEEE Trans. Parallel Distrib. Syst.}, vol.~7, no.~12, pp. 1281--1291,
  1996.

\bibitem{bcube}
C.~Guo, G.~Lu, D.~Li, H.~Wu, X.~Zhang, Y.~Shi, C.~Tian, Y.~Zhang, and S.~Lu,
  ``Bcube: a high performance, server-centric network architecture for modular
  data centers,'' in \emph{SIGCOMM}, P.~Rodriguez, E.~W. Biersack,
  K.~Papagiannaki, and L.~Rizzo, Eds.\hskip 1em plus 0.5em minus 0.4em\relax
  ACM, 2009, pp. 63--74.

\bibitem{dcell}
C.~Guo, H.~Wu, K.~Tan, L.~Shi, Y.~Zhang, and S.~Lu, ``Dcell: a scalable and
  fault-tolerant network structure for data centers,'' in \emph{SIGCOMM},
  V.~Bahl, D.~Wetherall, S.~Savage, and I.~Stoica, Eds.\hskip 1em plus 0.5em
  minus 0.4em\relax ACM, 2008, pp. 75--86.

\bibitem{odi}
D.~N. Jayasimha, B.~Zafar, and Y.~Hoskote, ``On chip interconnection networks
  why they are different and how to compare them,'' \emph{Intel}, 2006.

\bibitem{leighton}
F.~T. Leighton, \emph{Introduction to parallel algorithms and architectures:
  array, trees, hypercubes}.\hskip 1em plus 0.5em minus 0.4em\relax San
  Francisco, CA, USA: Morgan Kaufmann Publishers Inc., 1992.

\bibitem{empirical}
\BIBentryALTinterwordspacing
M.~Mirza-Aghatabar, S.~Koohi, S.~Hessabi, and M.~Pedram, ``An empirical
  investigation of mesh and torus noc topologies under different routing
  algorithms and traffic models,'' in \emph{Proceedings of the 10th Euromicro
  Conference on Digital System Design Architectures, Methods and Tools}.\hskip
  1em plus 0.5em minus 0.4em\relax Washington, DC, USA: IEEE Computer Society,
  2007, pp. 19--26. [Online]. Available:
  \url{http://dl.acm.org/citation.cfm?id=1302494.1302781}
\BIBentrySTDinterwordspacing

\bibitem{nakano}
K.~Nakano, ``Linear layout of generalized hypercubes,'' \emph{Int. J. Found.
  Comput. Sci.}, vol.~14, no.~1, pp. 137--156, 2003.

\bibitem{buses}
\BIBentryALTinterwordspacing
Y.~Pan, S.~Q. Zheng, K.~Li, and H.~Shen, ``An improved generalization of
  mesh-connected computers with multiple buses,'' \emph{IEEE Trans. Parallel
  Distrib. Syst.}, vol.~12, pp. 293--305, March 2001. [Online]. Available:
  \url{http://dx.doi.org/10.1109/71.914773}
\BIBentrySTDinterwordspacing

\bibitem{rolim95}
J.~D.~P. Rolim, O.~S{\'y}kora, and I.~Vrto, ``Optimal cutwidths and bisection
  widths of 2- and 3-dimensional meshes,'' in \emph{WG}, ser. Lecture Notes in
  Computer Science, M.~Nagl, Ed., vol. 1017.\hskip 1em plus 0.5em minus
  0.4em\relax Springer, 1995, pp. 252--264.

\bibitem{NOCsurvey}
E.~Salminen, A.~Kulmala, and T.~D. H, ``Survey of network-on-chip proposals,''
  \emph{Simulation}, no. March, pp. 1--13, 2008.

\bibitem{Youssef91}
A.~Youssef, ``Cartesian product networks,'' in \emph{ICPP (1)}, 1991, pp.
  684--685.

\bibitem{Youssef95}
------, ``Design and analysis of product networks,'' in \emph{Proceedings of
  the Fifth Symposium on the Frontiers of Massively Parallel Computation
  (Frontiers'95)}.\hskip 1em plus 0.5em minus 0.4em\relax Washington, DC, USA:
  IEEE Computer Society, 1995, pp. 521--.

\bibitem{torusbased}
\BIBentryALTinterwordspacing
D.~Zydek and H.~Selvaraj, ``Fast and efficient processor allocation algorithm
  for torus-based chip multiprocessors,'' \emph{Comput. Electr. Eng.}, vol.~37,
  pp. 91--105, January 2011. [Online]. Available:
  \url{http://dx.doi.org/10.1016/j.compeleceng.2010.10.001}
\BIBentrySTDinterwordspacing

\end{thebibliography}

\end{document}